\documentclass{article}
\usepackage{authblk}
\usepackage{ae}
\usepackage[T1]{fontenc}
\usepackage[ansinew]{inputenc}
\usepackage{amsmath}
\usepackage{amssymb}
\usepackage{color}
\usepackage{graphicx}
\usepackage{longtable}
\usepackage{latexsym}
\usepackage{stmaryrd}
\usepackage{amsfonts}
\usepackage{amsbsy}
\usepackage{mathrsfs}
\usepackage{psfrag}
\usepackage{enumerate}
\usepackage{MnSymbol}
\usepackage{vmargin}
\setmarginsrb{3cm}{2cm}{3cm}{2cm}{1cm}{1cm}{1cm}{1cm}
\usepackage{hyperref}
\hypersetup{colorlinks=false}

\usepackage{amsmath,amssymb,calc,amsfonts}
\usepackage{latexsym}

\usepackage{graphicx,calc,epsfig}
\def\ut#1{\rlap{\lower1ex\hbox{$\sim$}}#1{}}
\def\sdpb#1{\rlap{\lower1.5ex\hbox{$\Leftarrow$}}{#1}}

\newcommand{\C}{\mathbb{C}}

\newcommand{\be}{\nopagebreak[3]\begin{equation}}
\newcommand{\ee}{\end{equation}}
\newcommand{\bee}{\nopagebreak[3]\begin{equation*}}
\newcommand{\eee}{\end{equation*}}
\newcommand{\ba}{\nopagebreak[3]\begin{eqnarray}}
\newcommand{\ea}{\end{eqnarray}}
\newcommand{\baa}{\nopagebreak[3]\begin{eqnarray*}}
\newcommand{\eaa}{\end{eqnarray*}}
\newcommand{\la}{\label}
\DeclareFontFamily{U}{rsfs}{}         
\DeclareFontShape{U}{rsfs}{m}{n}{<5> rsfs5 <6><7> rsfs7          %
  <8><9><10><10.95><12><14.4><17.28><20.74><24.88> rsfs10}{}     %
\DeclareMathAlphabet{\mathfs}{U}{rsfs}{m}{n}                     %
\newcommand{\mfs}[1]{\mathfs {#1}}                               %
\newcommand{\va}{\scriptscriptstyle}

\newcommand{\van}{\scriptstyle}

\newcommand{\n}{{\nonumber}}

\newcommand{\sH}{{\mfs H}}

\newcommand{\sN}{{\mfs N}}

\newcommand{\Hk}{{\sH}_{kin}}

\newcommand{\su}{\mathfrak{su}}

\newcommand{\tr}{\mathrm{tr}}

\DeclareMathOperator{\sign}{sign}

\begin{document}
\title{\bf Horizon entropy with loop quantum gravity methods}

\author[1]{Daniele Pranzetti\thanks{daniele.pranzetti@gravity.fau.de}}
\author[1]{Hanno Sahlmann\thanks{hanno.sahlmann@gravity.fau.de}}

\affil[1]{{\it Institute for Quantum Gravity}

University Friedrich Alexander University Erlangen-N\"urnberg (FAU), 

Staudtstrasse 7 / B2, 91058 Erlangen, Germany.}

\sloppy
\maketitle

\pagestyle{plain}

\begin{abstract}
We show that the spherically symmetric isolated horizon can be described in terms of an SU(2) connection and a su(2)-valued one-form, obeying certain constraints. The horizon symplectic structure is precisely the one of 3d gravity in a first order formulation. We quantize the horizon degrees of freedom in the framework of loop quantum gravity, with methods recently developed for 3d gravity with non-vanishing cosmological constant. Bulk excitations ending on the horizon act very similar to particles in 3d gravity. The Bekenstein-Hawking law is recovered in the limit of imaginary Barbero-Immirzi parameter. Alternative methods of quantization are also discussed. 
\end{abstract}
\sloppy
\maketitle
\newpage{}

\section{Introduction}\la{Intro}

The standard black hole entropy calculation in loop quantum gravity (LQG) strongly relies on the interplay with the Chern-Simons theory describing the horizon degrees of freedom (dof). The relevance of TQFT in non-perturbative quantum gravity when a boundary of finite area is present was first pointed out in \cite{Smolin}. The central role of Chern-Simons theory was then further established in \cite{ABCK} by means of the {\it isolated horizon} (IH) boundary conditions \cite{IH}, providing a local definition of an isolated black hole more general and physically relevant than the notion of event horizon. The $U(1)$ gauge fixing adopted in \cite{ABCK} has been more recently relaxed for all physically relevant kinds of black holes. This was systematically derived and developed in the sequence of papers \cite{ENP, ENPP1, PP1}, providing a fully $SU(2)$-invariant Chern-Simons description of isolated horizons boundary theory. This analysis provided the theoretical framework for  analytical \cite{GM, ENPP2} and numerical \cite{BarberoSU(2)} techniques developed for the counting of the number of boundary dof.  Along the lines of the original point of view of \cite{ABCK}, the leading term for the IH entropy has been shown to be in agreement with the Bekenstein-Hawking semiclassical formula \cite{Bekenstein}   for a fixed numerical value of the  Barbero-Immirzi parameter $\beta$ given by $\beta_0= 0.274067...$. See \cite{D-PP} for a review of these results. 

This unexpected central role of the Barbero-Immirzi (B-I) parameter in recovering a semiclassical result of QFT on a fixed geometry has recently motivated an alternative scenario. In \cite{Complex} it has been noted that, by taking an analytic continuation of the dimension of the $SU(2)$ Chern-Simons Hilbert space on a punctured 2-sphere (modeling a quantum IH) to $SL(2,\C)$ together with some assumptions on the spin representations, the semiclassical result could be recovered without the numerical restriction  $\beta=\beta_0$. Such analytic continuation was interpreted as the passage to an imaginary B-I parameter, and this choice is physically preferred due to the correct transformation of the Ashtekar self-dual connection \cite{Ash-con} under space-time diffeomorphisms \cite{Samuel}. In particular, in \cite{Temp} it was shown how local Lorentz invariance underlies the strict connection between the analytic continuation to $\beta=i$ and the thermality of the quantum IH. 

However, regardless of the status of $\beta$, the fundamental role played by Chern-Simons theory in the black hole entropy calculation is evident. In order to claim this as a full success of the LQG approach, it would be desirable to have a quantization of the IH boundary theory completely within the kinematical framework of the theory and be able to perform the counting without relying on the Verlinde formula for the Chern-Simons Hilbert space dimension.  Moreover, the standard coupling between bulk and boundary theories, requiring identification of certain structures of LQG and Chern-Simons theory, presents a number of ambiguities which affects the entropy calculation and are at the core of some of the still open issues. A more uniform treatment uniquely in terms of the LQG formalism, besides making the whole derivation more sound, can help to solve the latter and also provide further insight on the aspects of the calculation mentioned above. 

A first attempt along this direction was made in \cite{S}, where some structures of the quantum deformation $SL_q(2)$ of the $SU(2)$ group (with $q$ the deformation parameter), expected to be associated to the Chern-Simons theory, appeared; however, a clear Hilbert space structure was still lacking there. In this paper we proceed further on this route. 

More precisely, in Section \ref{sec:symplectic} we show how the IH conserved presymplectic form can be re-expressed in terms of first order gravity variables and list the boundary conditions that these have to satisfy. In Section \ref{sec:NCC} we show how the Ashtekar-Barbero connection on the IH becomes non-commutative and we introduce a second new non-commutative connection in order to be able to rely on techniques developed in \cite{NPP1, P1} in the context of 2+1 gravity with non-vanishing cosmological constant to quantize the boundary theory using LQG techniques. The quantization is carried out in Section \ref{sec:quantization}, where the IH quantum state is defined by regularizing point punctures with finite loops, as required by the extended nature of the LQG configuration variables and in analogy to the proposal of \cite{GP}; we then define the physical scalar product of the horizon theory, imposing the quantum version of the boundary conditions. In Section \ref{sec:entropy} we use the equivalence \cite{P1} between the Chern-Simons observables expectation values and the physical amplitudes of 2+1 canonical LQG to compute the number of  IH dof by means of the physical scalar product previously defined. We find that the degeneracy of the boundary quantum state satisfies the Bekenstein holographic bound for $\beta=i$, thus providing further evidence for the new perspective advocated above. 
In Section \ref{sec:alternative} an alternative quantization scheme closer in spirit to the approach of \cite{S} is presented, by developing a comparison with the context of 2+1 gravity coupled to point particles. 
Section \ref{sec:summary} contains a summary of our results.
In this paper we focus our attention on the spherically symmetric case. 

\section{Isolated Horizon Presymplectic Form}\la{sec:symplectic}

In order to express the conserved presymplectic form in terms of BF variables\footnote{A description of non-rotating isolated horizons in terms of symmetry reduced $SO(1,1)$ BF theory was used in \cite{BF}.}, let us recall first some useful relations following from the IH boundary conditions (see \cite{D-PP} for more details and definitions).  
The phase-space variables of gravity in the first order formalism are given by the 2-form densitized triad $\Sigma^i$ (with $i,j,k=1,2,3$ and $I,J=0,i$ internal $SL$(2,$\C$) indices) defined as:
\begin{equation}
\Sigma^{IJ}\equiv e^I\wedge e^J~~~~~~~\Sigma^i\equiv \epsilon^i\,_{jk}\Sigma^{jk}
\end{equation}
and the 1-form extrinsic curvature $K^i=\omega^{0i}$, where $\omega^{IJ}$ is the spin connection defined by $\omega^{IJ}_a\equiv e^{Ib} \nabla_a e^J_b$
and related to the metric through the relation
$g_{ab}=e^I_ae^J_b\eta_{IJ}$,
where $\eta_{IJ}={\rm diag}(-1,1,1,1)$.

In terms of these phase-space variables we can write the presymplectic form for gravity as:
\be\label{SF1}
\kappa\Omega(\delta_1,\delta_2)=\int_M \delta_{[1}\Sigma^i\wedge \delta_{2]} K_i\,,
\ee
where $\kappa=8\pi G$, $M$ is a Cauchy surface representing space and $\delta_1,\delta_2\in T_p\Gamma$, i.e. they are vectors in the tangent space to the phase-space $\Gamma$ at the point $p$. $\Gamma$ is an infinite-dimensional manifold whose points $p$ are given by solutions to the Einstein equations and are labeled by a pair $p=(\Sigma, K)$.

We now want to introduce the Ashtekar-Barbero variables defined through the introduction of the connection $A^i_a$:
\be
A^i_a=\Gamma^i_a+\beta K^i_a\,,
\ee
where $\Gamma^i=-\frac{1}{2}\epsilon^{ijk}\omega_{jk}$ and $\beta$ is the Barbero-Immirzi parameter. The connection $A^i$ is still conjugate to $\Sigma^i$ and in terms of it the presymplectic form (\ref{SF1}) takes the form:
\ba\label{SF2}
\kappa\Omega(\delta_1,\delta_2)&=&\frac{1}{\beta}\int_M \delta_{[1}\Sigma^i\wedge \delta_{2]} \beta K_i+
\frac{1}{\beta}\int_M \delta_{[1}\Sigma^i\wedge \delta_{2]} \Gamma_i-
\frac{1}{\beta}\int_M \delta_{[1}\Sigma^i\wedge \delta_{2]} \Gamma_i\n\\
&=&\frac{1}{\beta}\int_M \delta_{[1}\Sigma^i\wedge \delta_{2]} A_i-
\frac{1}{\beta}\int_{\partial M} \delta_{[1}e^i\wedge \delta_{2]} e_i\,,
\ea
where $\partial M$ is the boundary of $M$. If we assume $\partial M$ to correspond to a 2-sphere cross-section $IH$ of $M$ with an isolated horizon  $\Delta$, then the isolated horizon boundary conditions \cite{IH} imply  the following relation to hold on the 2-sphere:
\be
\sdpb{F}^i(A_+)=-\Psi_2\sdpb{\Sigma}^i
\ee
from which
\be\label{eq:IH}
F^i(\Gamma)=-Re(\Psi_2)\sdpb{\Sigma}^i+\frac{1}{2}\epsilon^i\,_{jk}\sdpb{K}^j\wedge \sdpb{K}^k\,,~~~~~~
d_\Gamma \sdpb{K}^i=-Im(\Psi_2)\sdpb{\Sigma}^i\,,
\ee
where $A_+^i=\Gamma^i+i K^i$, $\Psi_2$ is the only non-vanishing Weyl scalar, the curvature $F^i(A)$ is given by $F^i(A)=dA^i+\frac{1}{2}\epsilon_{ijk} A^j\wedge A^k$; the double arrows denote the pull-back to 2-sphere $ IH$ and we will omit them from now on to lighten the notation. In particular, in the spherically symmetric case ($\Psi_2=\frac{2\pi}{a_{\va IH}}$), the above conditions imply
\be
F^i(A)=-\frac{\pi }{a_{\va IH}}(1-\beta^2){\Sigma}^i\,,~~~~~~d_\Gamma {K}^i=0\la{dK}\,,
\ee
where $a_{\va IH}$ is the area of the isolated horizon.
In \cite{ENPP1}, by means of a special gauge where the tetrad $(e^I)$ is such that $e^1$ is normal 
to $IH$ and $e^2$ and $e^3$ are tangent to $IH$, it has been shown that \eqref{dK} implies
${K}^1=0$,
which in turn shows that
$v\righthalfcup {\Sigma}^1 \wedge {K}_1=0$, where $v$ is a vector field tangent to $IH$. Since in the chosen gauge the pull back of $\Sigma^2$ and $\Sigma^3$ on the horizon is zero, then one has
\be\la{Sigma-K}
v\righthalfcup {\Sigma}^i \wedge {K}_i=0\,.
\ee
Therefore, \eqref{Sigma-K} being true in a particular gauge is true in general, since it is a gauge invariant relation.
Another useful relation valid on $IH$ is \cite{ENPP1}
\be\label{eq:KK=Sigma}
{K^j}\wedge {K^k}\epsilon_{ijk}= \frac{2\pi }{a_{\va IH}} {\Sigma}^i\,.
\ee

The IH boundary conditions also restrict the variations
$\delta=(\delta\Sigma,\delta A)\in {\rm T_p} (\Gamma)$   such that
for fields pulled back  on the horizon they are given by  linear
combinations of $SU(2)$ internal gauge transformations and
diffeomorphisms which preserve the preferred foliation of $\Delta$.

In \cite{ENPP1} it has been shown that  the IH boundary conditions listed above preserve the presymplectic form \eqref{SF2}, in the sense that it is independent of $M$.
We are now going to show that the boundary term in \eqref{SF2} can be rewritten in terms of first order gravity variables on $IH$.
\vskip.2cm

\noindent{\bf Proposition:}  In terms of Ashtekar-Barbero connection and its conjugate momentum variables the conserved presymplectic structure of a spherically symmetric IH takes the form
\be\label{SF3}
\kappa\Omega(\delta_1,\delta_2)=\frac{1}{\beta}\int_M \delta_{[1}\Sigma^i\wedge \delta_{2]} A_i+
\frac{1}{\beta^2}\sqrt{\frac{a_{\va IH}}{2\pi}}\int_{\va IH} \delta_{[1}e^i\wedge \delta_{2]}A_i\,.
\ee

\vskip.4cm
\noindent {\it Proof:} we need to show that the phase space one-form $\Theta(\delta)$ defined by 
\be
\Theta(\delta) \equiv  \int_{\va IH}  e^i \wedge
\delta e_i +\frac{1}{\beta}\sqrt{\frac{a_{\va IH}}{2\pi}}\int_{\va IH}
e^i\wedge \delta A_i
\ee 
is closed, where the exterior derivative of $\Theta(\delta)$ is given by
\bee
{\frak d}\Theta_0 (\delta_1,\delta_2)=\delta_1(\Theta_0(\delta_{2}))-\delta_2 (\Theta_0(\delta_{1}))\,.
\eee
We saw above that the gauge symmetry transformations allowed by the IH boundary conditions on $IH$ are given by infinitesimal $SU(2)$ transformations and diffeomorphisms tangent  to the horizon. Therefore let us consider variations of the form $\delta=\delta_{\alpha}+\delta_v$, 
where $\alpha: IH\rightarrow su(2)$ and $v$ is a vector field tangent to $IH$. Under such transformations we have 
\baa
&&\delta_{\alpha}e^i=[\alpha, e]^i\,,~~~
\delta_{\alpha}A^i=-d_A \alpha^i\,,\\
&&\delta_v e^i=\mathscr{L}_v e^i
=v\righthalfcup de^i + d(v\righthalfcup e^i)=(\delta^*_v-\delta_{\alpha(A,v)})e^i=v\righthalfcup d_A e^i+d_A(v\righthalfcup e^i)-[v\righthalfcup A,e]^i\,,\\
&&\delta_v A^i=\mathscr{L}_v A^i=(\delta^*_v-\delta_{\alpha(A,v)})A^i=v\righthalfcup F^i(A) + d_A(v\righthalfcup A^i)\,,
\eaa
where $\alpha(A,v)=v \righthalfcup A$ and $\delta^*_v$ is defined as 
$\delta^*_v  A^i=v\righthalfcup F^i(A)$ and  $ \delta^*_v e^i=v\righthalfcup d_A e^i+d_A(v\righthalfcup e^i)$.

Let us also derive a useful relation, which will represent an extra boundary condition due to the doubling of the boundary d.o.f. introduced with the new boundary term in \eqref{SF3}, namely
\be\label{dAe}
d_{A}e^i=d_\Gamma e^i+\beta\epsilon^i\,_{jk}K^j\wedge e^k=\beta\epsilon^i\,_{jk}K^j\wedge e^k
=-\beta\sqrt{\frac{2\pi}{a_{\va IH}}}\Sigma^i\,,
\ee
where in the second passage we have used the Cartan equation $de^i+\epsilon^i\,_{jk}\Gamma^j\wedge e^k=0$ and in the last one the relation 
\be\la{KSigma}
K^i_a=-\sqrt{\frac{2\pi}{a_{\va IH}}} e^i_a
\ee
derived in \cite{ENPP1} (from which \eqref{eq:KK=Sigma} follows).
We also recall that on a 2-manifold
$A\wedge v\righthalfcup B=- v\righthalfcup A\wedge B$
for any 2-form $A$ and 1-form $B$, while
$A\wedge v\righthalfcup B= v\righthalfcup A\wedge B$
for any 1-form $A$ and 2-form $B$.

Let us start with the gauge transformations:
\baa
{\frak d}\Theta (\delta,\delta_\alpha)
&=&\int_{\va IH}2\delta_{[}e^i\wedge \delta_{\alpha]}e_i
+\frac{1}{\beta}\sqrt{\frac{a_{\va IH}}{2\pi}}\int_{\va IH}\left( \delta_{[}e^i\wedge \delta_{\alpha]}A_i
- \delta_{[\alpha}e^i\wedge \delta_{]}A_i\right)\n\\
&=&\int_{\va IH}4\delta e^i\wedge \epsilon_{ijk}\alpha^je^k
-\frac{2}{\beta}\sqrt{\frac{a_{\va IH}}{2\pi}}\int_{\va IH}\left( \delta e^i\wedge d_A \alpha_i
+ \epsilon_{ijk}\alpha^j e^k\wedge \delta A^i\right)\n\\
&=&-2\int_{\va IH} \delta(\Sigma^i+\frac{1}{\beta}\sqrt{\frac{a_{\va IH}}{2\pi}}d_A e^i)\alpha_i=0\,,
\eaa
where in the last passage we used \eqref{dAe}. For diffeomorphisms we have:
\baa
{\frak d}\Theta (\delta,\delta_v)
&=&\int_{\va IH}4\delta e^i\wedge \delta_v e_i
+\frac{2}{\beta}\sqrt{\frac{a_{\va IH}}{2\pi}}\int_{\va IH}\left( \delta e_i\wedge(\delta^*_v-\delta_{\alpha(A,v)})A^i)-(\delta^*_v-\delta_{\alpha(A,v)})e_i\wedge \delta A^i \right)\n\\
&=&4\int_{\va IH}\!\!\!\delta\left(de^i\wedge v\righthalfcup e^i\right)
+{\frac{2}{\beta}\sqrt{\frac{a_{\va IH}}{2\pi}}\int_{\va IH}\!\!\!\delta \left(  e_i \wedge v\righthalfcup F^i(A)\right)}
+\frac{2}{\beta}\sqrt{\frac{a_{\va IH}}{2\pi}}\int_{\va IH}\!\!\!\delta \left(  d_A e^i v \righthalfcup A_i\right)\n\\
&=& {-4\int_{\va IH}\!\!\!\delta( \epsilon^i\,_{jk}\Gamma^j\wedge e^k \wedge v\righthalfcup e_i)}
-{\frac{(1-\beta^2)}{\beta}\sqrt{\frac{2\pi}{a_{\va IH}}}\int_{\va IH}\!\!\!\delta \left(  e_i \wedge v\righthalfcup \Sigma^i\right)}
-2\int_{\va IH}\!\!\!\delta \left(  \Sigma^i v \righthalfcup A_i\right)\n\\
&=&-\beta\int_{\va IH}\!\!\!\delta( K^i \wedge v\righthalfcup\Sigma_i)=0\,,
\eaa
where in the third line we have used the result of the previous calculation with $\alpha=v \righthalfcup A$ and the relation $\delta d_A A^i=\delta F^i(A))$, in the fourth Cartan's equation, and eq. \eqref{Sigma-K} for the vanishing of $e_i \wedge v\righthalfcup \Sigma^i$ in the last line. $\square$
\vskip.5cm

Hence, the IH conserved presymplectic form can be expressed in the form \eqref{SF3}, which shows how the boundary theory can be parametrized by the variables $(A, e)$ satisfying the boundary conditions
\ba
&&F^i(A)=-\frac{\pi }{a_{\va IH}}(1-\beta^2)\, {\Sigma}^i\la{BC1}\\
&&d_A e^i=-\beta\sqrt{\frac{2\pi}{a_{\va IH}}}\,\Sigma^i\la{BC2}\,.
\ea

\section{Non-commutative connection}\la{sec:NCC}
On the isolated horizon $IH$ we have a $2+1$ theory. In the previous section we have seen that, upon the standard 2+1 decomposition, the phase space of the theory can be parametrized
by the pullback to $IH$ of the Ashtekar-Barbero connection and the triad. In local coordinates
we can express them in terms of the 2-dimensional connection $A_{a}^{i}$
and the dyad field $e_{a}^{i}$ where $a=1,2$ are space coordinate 
indices on $IH$ and $i,j=1,2,3$ are $\su(2)$ indices. 
As already pointed put in \cite{ENPP1}, the boundary term in the presymplectic from \eqref{SF2} implies that the horizon dyad field satisfy the Poisson bracket
\be \la{ee}
\{e^i_a(x), e^j_b(y)\}=-\kappa\beta \epsilon_{ab}\ \delta^{ij}\delta^{(2)}\left(x,\, y\right)\,,
\ee
where $\epsilon_{ab}$ is the 2d Levi-Civita tensor. 
At the same time, if we parametrize the IH phase space in terms of first order gravity variables $(A,e)$, the boundary term in the presymplectic from \eqref{SF3} indicates that the Poisson bracket
among them is given by 
\begin{equation}\label{Ae}
\{ A_{a}^{i}\left(x\right),\, \tilde e_{b}^{j}\left(y\right)\} =\kappa\beta\epsilon_{ab}\ \delta^{ij}\delta^{(2)}\left(x,\, y\right)
\end{equation}
where 
 \be
\tilde e^i_a := \frac{1}{\beta}\sqrt{\frac{a_{\va IH}}{2\pi}}e^i_a\,.
 \ee
 The two Poisson brackets \eqref{ee} and \eqref{Ae} are consistent with each other as soon as we take into account the relation \eqref{KSigma} holding on the horizon 2-sphere. In particular, this implies that the Ashtekar-Barbero boundary connection becomes non-commutative. This should not be surprising if one wants, as standardly done in the literature, interpret the boundary condition \eqref{BC1} as the eom of the $SU(2)$ Chern-Simons theory on a punctured 2-sphere, since the Chern-Simons connection in the l.h.s. of \eqref{BC1} is non-commutative. Despite such non-commutativity, the theory \eqref{Ae}, \eqref{BC1}, \eqref{BC2}
 bears a strong resemblance with 2+1 gravity with cosmological constant\footnote{in Section \ref{sec:alternative} we will present an alternative point of view where the horizon theory is treated as genuine BF 2+1 gravity coupled to point particles. }.
Let us clarify  this classical set-up so that it will then be straightforward to apply LQG techniques developed in that context to quantize the boundary theory. In order to do so, we introduce a new connection
\be
\tilde A^i_{a}=A^i_a+\alpha( a_{\va IH}) \tilde e^i_a\,,
\ee
where $\alpha(a_{\va IH})$ is a function of the IH area $a_{\va IH}$ to be determined by
expressing the condition \eqref{BC1} as a flatness condition for the new connection. More precisely,
\bee
F^i(\tilde A)=d\tilde A^i+\frac{1}{2}\epsilon^i\!\,_{lm} \tilde A^l \wedge \tilde A^m=F^i(A)+  (\frac{ a_{\va IH}}{2\pi \beta^2}\frac{\alpha^2}{2}-\alpha) \Sigma^i=0\,,
\eee
where in he last passage we used the boundary condition \eqref{BC2}. Therefore, the condition \eqref{BC1} is recovered once
$\alpha_{\pm}=\beta(\beta\pm1)2\pi/a_{\va IH}$.
The IH boundary conditions can thus be re-expressed as
\ba
&&F^i(\tilde A)=0\la{BC3}\\
&&
d_A \tilde e^i=- \Sigma^i\,, \la{BC4}
\ea
where
\ba
&&A^i_a=\Gamma^i_a+\beta K^i_a=\Gamma^i_a-\frac{2\pi\beta^2}{a_{\va IH}}\tilde e^i_a\la{A}=\Gamma^i_a-\frac{\beta}{2 \ell^2_{\va P}(1-\beta^2) k}\tilde e^i_a\la{A}\,,\\
&&\tilde A^i_{a}=A^i_a+\alpha_\pm \tilde e^i_a=\Gamma^i_a\pm\frac{2\pi\beta}{a_{\va IH}}\tilde e^i_a\la{Ak}=\Gamma^i_a\pm\frac{1}{2 \ell^2_{\va P}(1-\beta^2) k}\tilde e^i_a\la{Ak}\,,
\ea
and we have used the relation \cite{ENPP1}
\be
\label{eq:k}
k=\frac{a_{\va IH}}{4\pi \ell^2_{\va P}\beta(1-\beta^2)}
\ee
between the Chern-Simons level $k$ and the IH area $a_{\va IH}$.

The boundary condition \eqref{BC3} imposes the flatness of the non-commutative connection \eqref{Ak}, in analogy to the treatment of \cite{NPP1, P1} for 2+1 gravity in presence of a non-vanishing cosmological constant. While \eqref{BC4} encodes a modification of the Gauss constraint encoding singularities in the torsion of the Ashtekar-Barbero connection on the boundary in the form of punctures induced (in the quantum theory) from the bulk spin network links piercing $IH$. This is analog to the case of 2+1 gravity coupled to point particles  \cite{NP2} . We can thus combine the LQG techniques developed in the framework of 2+1 gravity to quantize the boundary theory on the IH.

\section{Quantization}\la{sec:quantization}

We now want to quantize the IH boundary theory parametrized by the BF variables $(A^i, \tilde e_i)$ and satisfying the constraints \eqref{BC3}, \eqref{BC4} just relying on LQG techniques. Then, one can  think to extend the quantization techniques of the bulk to the isolated horizon. Therefore,  the basic kinematical observables on the horizon are given by the holonomy of the
connection and appropriately smeared functionals of the dyad 
field $\tilde e$. Namely, one can find an irreducible representation of the quantum counterpart of these observables on a kinematical Hilbert space $\Hk^{IH}$ whose states are given by functionals $\Psi[A]$ of the
(generalized) connection $A$ which are square-integrable with
respect to a diff-invariant measure. 

The non-commutativity of the Ashtekar-Barbero connection on the IH 2-sphere does not represent an obstacle to the construction of the IH Hilbert space. This is the case since the non-commutative holonomy acting on the Ashtekar-Lewandowski vacuum \cite{AL} still has a multiplicative action. Moreover, the intersection of two such holonomies has been explicitly computed in \cite{NPP1} and shown to reproduce the Kauffman's crossing bracket \cite{KL}; in particular, the action of one non-commutative holonomy on another can again be recast in a multiplicative form. This allows us to apply standard LQG kinematical techniques to the construction of the IH Hilbert space. 
Therefore, holonomies of $A$ are quantized as in the 3+1 theory on the  Hilbert space $L_2(\overline{\mathcal{A}^{(2)}},\text{d}\mu_{\text{AL}})$ via multiplication\footnote{As explained below, the restricted set of boundary observables that will be relevant for the entropy calculation are formed only by loops around each puncture which do not intersect each other together with a set of holonomies defined on paths connecting each loop to a same single point. It is for this second set of holonomies that one should use the Kauffman bracket to represent the action of the non-commutative connection in a multiplicative form. However, the physical scalar product of the isolated horizon theory can be defined (see below) such that this set of holonomies plays no effective role and the boundary observables become to all purposes commutative. }, while 
\begin{equation}\la{flux}
\tilde e^i(\eta)=\int_\eta \tilde e^i_a \dot\eta^a
\end{equation}
is the analog of the 3d flux, in the sense that the quantity \eqref{flux} represents the flux of $e$ across the
one-dimensional paths $\eta^a(t)\in IH$, with $ \dot\eta^a=d\eta^a/dt$.
It is quantized analogously such that the associated operator acts
non-trivially only on holonomies $h_{\gamma}$ along a path $\gamma\in
IH$ that are transversal to $\eta$, namely
\begin{equation}
[\hat {\tilde e}(\eta), h_\gamma]=i\hbar\kappa\beta \sum_{p\in{\eta\cap\gamma}} \sign(\epsilon_{ab}\dot{\eta}^a\dot{\gamma}^b({p})) h_{\gamma_2({p})}J_i h_{\gamma_1({p})}\,,
\end{equation}
i.e. acts as the derivative operator $\hat {\tilde e}_{a}^{i}=-\mathrm{i}\hbar\kappa\beta\epsilon_{ab}\ \delta_{j}^{i}{\delta}/{\delta A_{b}^{j}}\,$.

\begin{figure}[ht]
\centering
\includegraphics[scale=0.4]{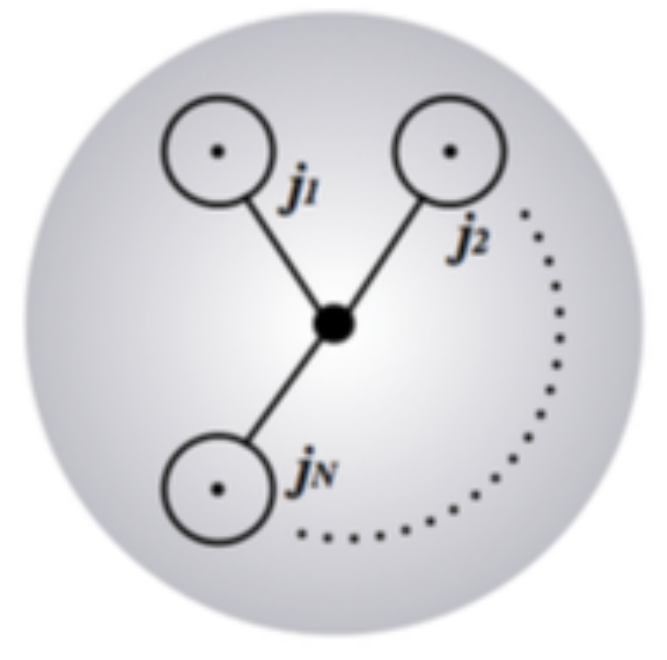}
\caption{States of the quantum isolated horizon.}
\label{fig:states}
\end{figure}

In order to impose the curvature constraint, we are going to use its non-commutative connection formulation \eqref{BC3} so to be able to import techniques developed in \cite{NPP1, P1}. But let us first concentrate on the modified Gauss law  \eqref{BC4}. From the LQG quantization of the densitized triad $\Sigma^i$ in the bulk we have
 \begin{equation}
\label{gammasigma} \epsilon^{ab}\hat{\Sigma}^i_{ab}(x) = 2\kappa
\beta \sum_{p \in \Gamma\cap IH} \delta(x,x_p) \hat{J}^i(p)\,,
\end{equation}
where the fixed graph $\Gamma \subset M$ has end points on $IH$ denoted $\Gamma\cap IH$ and the $\hat J$s satisfy the $\su(2)$ algebra  $[\hat{J}^i(p),\hat{J}^j(p)]=\epsilon^{ij}_{\ \ k} \hat{J}^k(p)$. 
Therefore, the bulk geometry induces conical singularities in the boundary torsion, which can be interpreted as point particles. It follows that, in order to remove ambiguity in the definition of the boundary connection at the location of the point particles, these  have to be blown-up to circles  \cite{NP2}.  These new boundaries on the horizon then inherit the spin-$j$ irrep carried by the corresponding bulk link piercing the horizon. In this way, quantum IH states can then be represented by a collection of small loops $\ell_i$ ($i=1,...,N$, $N$ being the total number of particles) colored with $SU(2)$ irreps $j_i$, each surrounding one puncture and connected by links forming a single intertwiner\footnote{This last property of the IH states is a consequence of a global constraint that follows from \eqref{BC1}, namely that the holonomy around a contractible loop encircling all particles be trivial.}  as in  Figure \ref{fig:states}. 

With this regularization, the modified Gauss law \eqref{BC4} can be seen as a relation between the flux of the horizon electric field $\tilde e$ across a given circle $\ell_i$ and the flux of the bulk electric field $\Sigma$ across the surface encircled by $\ell_i$. The imposition of such relation can then be implemented as in \cite{NP2} by associating an intertwiner $\iota_{i}$ of $SU(2)$ to each boundary $\ell_i$; note however that, differently from the case of \cite{NP2}, there is no free magnetic number associated to each particle link, since these are now connected to the rest of the bulk graph. Given the restricted structure of the horizon states depicted in FIG. \ref{fig:states}, each  $\iota_{i}$ is a trivial bivalent intertwiner.

We now analyze the imposition of the curvature constraint \eqref{BC1}. We see that away from the particles, the (non-commutative) connection $A^i$ is flat. Around each particle, i.e. along each loop, the curvature picks up a contribution proportional to the flux of the $\Sigma^i$ field. We saw in Section \ref{sec:NCC} that this curvature constraint at each puncture can again be expressed in terms of the flatness condition of a new non-commutative connection \eqref{Ak}. This allows us to use the analysis of \cite{P1, NP1} to define a projector into the physical Hilbert space of the boundary theory. In fact, if we introduce a cellular decomposition $\Delta_{IH}$
of the horizon 2-sphere $IH$---with plaquettes $p\in\Delta_{IH}$ of coordinate area smaller or equal to $\epsilon^{2}$---,  the curvature constraint can be  written as
\be\la{CN}
C[N]=\lim_{\epsilon\rightarrow0}\sum_{p\nin \cup \ell_i}\tr\left[N_{p}\,
W_{p}\left(A\right)\right]+\lim_{\epsilon\rightarrow0}\sum_{p\in \cup \ell_i}\tr\left[N_{p}\,
W_{p}\left(\tilde A\right)\right]=0
\ee
where $W_{p}=1+\epsilon^{2}F+o(\epsilon^{2})\in SU(2)$ is
the Wilson loop of the connection $A$, $\tilde A$ in the spin-1/2 representation. 
It is immediate to see that the only non-vanishing contributions to the commutator of the constraint \eqref{CN} with itself, when acting on a gauge invariant state, come from the commutator of any of the terms $p$ with itself, i.e. of the form $\left[\tr\left[N_{p}\, W_{p}\left(A\right)\right], \tr\left[M_{p}\,W_{p}\left(A\right)\right]\right]$ or $\left[\tr\left[N_{p}\, W_{p}\left(\tilde A\right)\right], \tr\left[M_{p}\,W_{p}\left(\tilde A\right)\right]\right]$. In \cite{P1}, by means of techniques developed in \cite{NPP1, PP2}, it has been shown that such commutators are anomaly-free if and only if the infinitesimal loop evaluates to the quantum dimension, namely 
\be\la{q-dim-j}
\begin{array}{c}\includegraphics[width=.8cm]{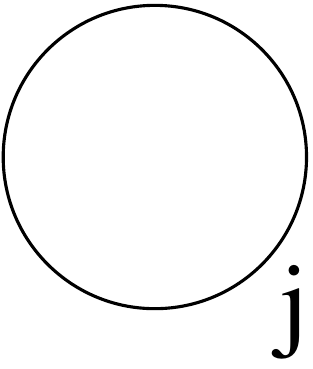}\end{array}=(-)^{2j}[2j+1]_q=(-)^{2j}\frac{q^{2j+1}-q^{-(2j+1)}}{q-q^{-1}}\,,
\ee
where now 
\be\label{q} q=  \left\{ \begin{array}{ll}
         e^{\frac{2\pi i\beta^2 }{a_{\va IH}}\frac{\kappa \beta \hbar}{2} }=e^{\frac{2\pi i }{k}\frac{ \beta^2 }{(1-\beta^2)} }, ~~{\rm for}~ p\nin \cup \ell_i\\
        e^{\frac{2\pi i\beta }{a_{\va IH}}\frac{\kappa \beta\hbar}{2} }=e^{\frac{2\pi i }{k}\frac{ \beta }{(1-\beta^2)} },~~{\rm for}~ p\in \cup \ell_i\end{array} \right. \ee
as follows from the expression \eqref{A} and \eqref{Ak}.
Therefore, the condition \eqref{q-dim-j} implies that, at each plaquette, the recoupling theory of the classical $SU(2)$ group has to be replaced with the one of the quantum group $U_q SL(2)$; however, whether the plaquette surrounds a particle or not the deformation parameter $q$ takes one of the two different expressions above. 
We thus have two quantum group recoupling theories entering the quantization of the curvature constraint on the horizon.
This suggests that, following the construction of \cite{P1, NP1}, the physical scalar product for the IH boundary theory between two horizon states $s, s'$  can be written as
\be\la{phys}
\langle s, \;s^{\prime}\rangle_{\va phys}=\langle P[A, \tilde A] s,\;s^{\prime} \rangle\,,
\ee
where
\ba\la{projector}
P[A, \tilde A]&=&\lim_{\epsilon\rightarrow 0} \ \  \prod_{p\nin \cup \ell_i} \delta(W_{p}(A)) \prod_{p\in \cup \ell_i} \delta(W_{p}(\tilde A))\n\\
&=& \lim_{\epsilon\rightarrow 0} \ \  \sum_{j_{\va p}}\
   \prod_{p\nin \cup \ell_i}(-)^{2j_{\van p}}[2j_{\van p}+1]_q\
 \chi_{\va j_p}(W_{p}(A))\prod_{p\in \cup \ell_i}(-)^{2j_{\van p}} [2j_{\van p}+1]_q\ \chi_{\va j_p}(W_{p}(\tilde A))
 \ea
is the projector operator into the physical Hilbert space of the IH boundary theory. In the last line of the expression above the deformation parameter $q$ at each plaquette $p$ takes either one of the two different values \eqref{q} according to the presence or not of a puncture inside $p$.


\section{Entropy}\la{sec:entropy}

In order to compute the microcanonical BH entropy we need to derive the dimension of the physical Hilbert space of the IH boundary theory and then take its logarithm, according to the standard relation $S=\log {(\sN)}$, with $\sN$ the number of horizon micro-states compatible with the given macroscopic horizon area $a_{\va IH}$. The quantity $\sN$ can now be obtained from the relation between the Chern-Simons partition function on a  three manifold containing a collection of unlinked, unknotted Wilson lines and the scalar product between states of the associated Hilbert space used by Witten in his approach to the Jones polynomial \cite{W}.

More precisely, given a three manifold $M$ obtained from the connected sum of two three manifolds $M_1$ and $M_2$ joined along a two sphere $S^2$ and containing $N$ unlinked and unknotted circles $C_i$ with $SU(2)$ irreps $j_i$ associated to them, we denote the corresponding Chern-Simons partition function (or Feynman path integral) as $Z(M; \prod_{i=1}^N C_i)$; then
\be\la{Z}
Z(M; \prod_{i=1}^N C_i)=\langle \Psi_2|\Psi_1\rangle\,,
\ee
where $\Psi_1$ is the vector determined by the Feynman path integral on $M_1$ in the physical Hilbert space associated to the Riemann surface $S^2$  and $\Psi_2$ the vector  determined by the Feynman path integral on $M_2$ in the dual Hilbert space; $\langle \cdot|\cdot\rangle$ indicate the physical scalar product on this Hilbert space.
The expression \eqref{Z} correspond to the unnormalized expectation value of the link formed by the collection of circles $C_i$ from which Jones knot invariants can be derived \cite{W}. In the case $M=S^2\times S^1$, with $S^1$ corresponding to a compact time direction, we have 
\be\la{dim}
Z(S^2\times S^1; \prod_{i=1}^N C_i)= {\rm dim} \sH_{S^2; \otimes_i j_i}\,,
\ee
where the r.h.s. corresponds to the dimension of the Hilbert space on a punctured two sphere. The Reshetikhin-Turaev-Witten (RTW) invariant of a closed 3-manifold \cite{RT} provides a precise definition of the Chern-Simons path integral \eqref{Z}; at the same time, the Turaev-Viro (TV) invariant \cite{TV}, which represent a state-sum model for 3-dimensional Euclidean quantum gravity with positive cosmological constant $\Lambda$ \cite{TV-grav}, has been shown to be related to the Witten's Chern-Simons TQFT \cite{W} by the theorem $Z_{TV}(M)=|Z_{WRT}(M)|^2$  \cite{Turaev}. From a LQG perspective, it has been shown in \cite{P1} that the Turaev-Viro amplitudes can be recovered from the physical scalar product of the 2+1 theory with $\Lambda>0$ using the same formalism introduced in the previous section. In particular, with a proper relation between $\Lambda$ and $k$ (or $a_{\va IH}$),  the equivalent of the physical scalar product \eqref{phys}, \eqref{projector} provides an explicit definition of the r.h.s of \eqref{Z}, allowing us to recover the link expectation values computed via the Chern-Simons partition function.

Therefore, we can now use these results together with the relation \eqref{dim} to compute
the number $\sN$ of IH quantum states by means of the physical scalar product \eqref{phys} of the boundary Hilbert space. 
Following this logic, we thus have
\ba\la{phys1}
\sN=< P \emptyset,\!\!\begin{array}{c}  \includegraphics[width=1.5cm]{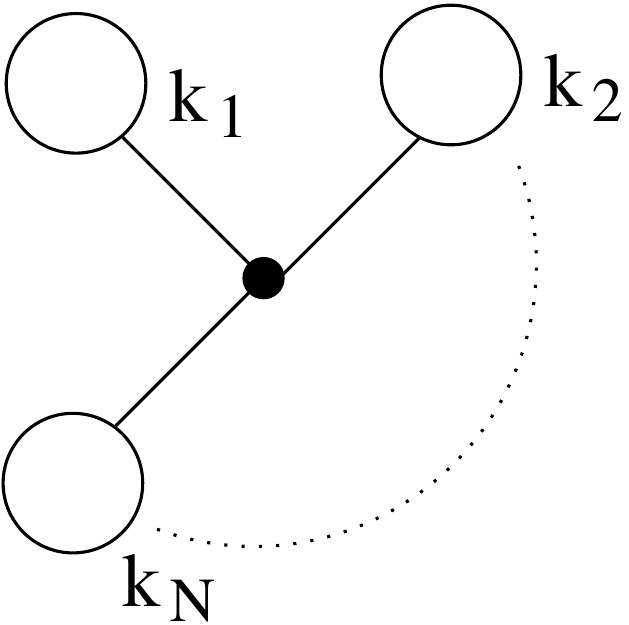}\end{array}\!\! >&=&
\lim_{\epsilon\rightarrow 0}\int  \left(\prod_h dg_{\va h}\right) \prod_i\chi_{\va k_i}(g_{\va \ell_i})
\
   \prod_{p\nin \cup \ell_i}\ \sum_{j_{\va p}}\ (-)^{2j_{\van p}}[2j_{\van p}+1]_q\ 
 \chi_{\va j_p}(W_{p}(A))\n\\
 &\times&\prod_{p\in \cup \ell_i}\ \sum_{j_{\va p}}\ (-)^{2j_{\van p}} [2j_{\van p}+1]_q\ \chi_{\va j_p}(W_{p}(\tilde A))
\ea
where $g_{\va \ell_i}$ is the holonomy along the  loop going around the $i$-th particle in the IH state and $dg_{\va h}$ corresponds to the invariant $SU(2)$-Haar measure. In relation to the notation in \eqref{Z}, we have identified the state $|\Psi_1\rangle$ with the isolated horizon state depicted in FIG. \ref{fig:states} and $|\Psi_2\rangle$ with the vacuum state; one can always find a decomposition of $M$ such that this is the case and the final result is insensitive to such choice  (the same expression for the r.h.s. of \eqref{phys1} would be obtained for any other decomposition).
 A graphical representation of the physical scalar product \eqref{phys1} is depicted in Figure 2.
\begin{figure}[h!]\la{fig:scalar}
\centering
  \includegraphics[width=6.1cm]{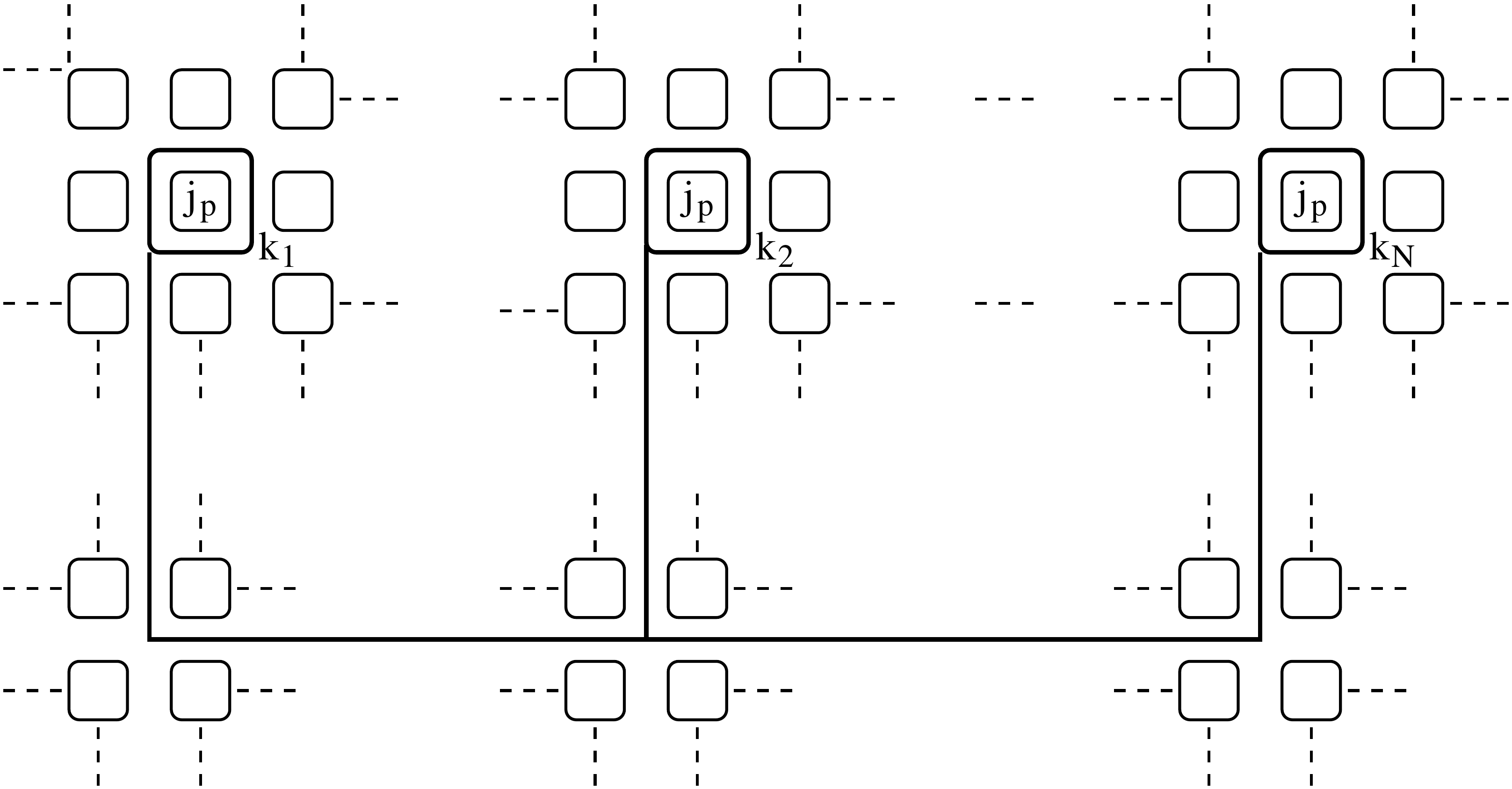}
   \caption{Graph structure of the projector operator and the quantum IH state inside the physical scalar product \eqref{phys1}.}
\end{figure}

In order to proceed with the evaluation of the physical scalar product, let us recall that, due to the (discrete) Bianchi identity, there is a redundancy in the  in the product of delta distributions entering the expression of the projection operator and associated to the plaquettes regulating the two sphere horizon surface. A way to deal with such redundancy consists of eliminating the holonomy $W_{p}(A)$ around an arbitrary plaquette $p\nin \cup \ell_i$. By doing so, it is immediate to see that when we perform the group integration over the edges belonging to $p\nin \cup \ell_i$ the intertwiner structure disappears from the scalar product (all the links connecting the loops $\ell_i$ are forced into the $j=0$ irrep). This is how the disappearance of the intertwiner structure mentioned above takes place and the evaluation of \eqref{phys1} is considerably simplified. As it is well known (see, e.g., \cite{ENPP2}), ignoring the intertwiner structure  affects the logarithmic correction to the entropy result, but does not modify the leading term. Thus, for the purposes of this paper, such simplification is irrelevant and our entropy result should be compared to the large $k$ limit of the standard counting one can find in the literature. 
We, hence, end up with 
\be \la{phys2}
\sN= \lim_{\epsilon\rightarrow 0}\int  \left(\prod_h dg_{\va h}\right) \prod_i  \prod_{p}\ \sum_{j_{\va p}}\ (-)^{2j_{\van p}} [2j_{\van p}+1]_q\ \begin{array}{c}  \includegraphics[width=1.2cm]{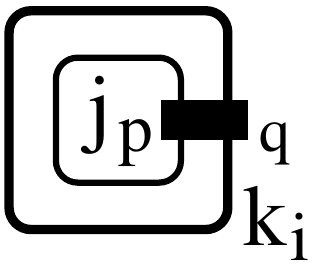}\end{array}\!\!\!= \prod_i (-)^{2k_{\van i}} [2k_{\van i}+1]_q
= \prod_i e^{2\pi i k_{\van i}} [2k_{\van i}+1]_q\,,
\ee
where, as shown in \cite{P1}, the $q$-box integration 
has to be  performed according to the renormalized skein relation
\bee\label{integration}
\begin{array}{c}  \includegraphics[width=.95cm,angle=360]{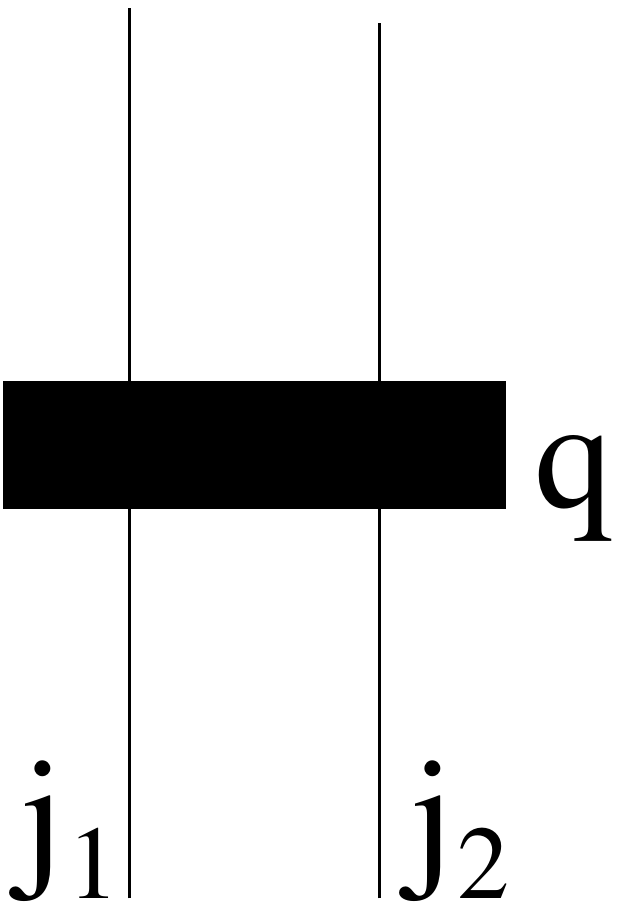}\end{array}=\frac{1}{(-)^{2j_1} [2j_1+1]_q}\delta_{j_1 j_2}\begin{array}{c}  \includegraphics[width=.8cm,angle=360]{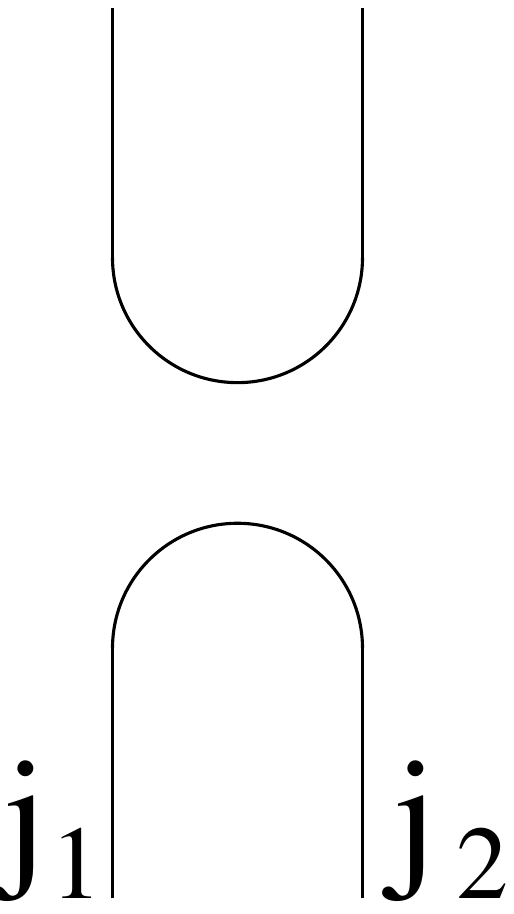}\,.\end{array}
\eee

Therefore, we see that at each puncture, besides the usual term given by the $SU(2)$ irreducible representation dimension associated to it (in the large $k$ limit), a new degeneracy factor appears which reproduces the Bekenstein's holographic bound for $\beta=i$, namely 
\be\la{deg}
\exp{(2\pi i k_i)}=\exp{(a_i/4\ell^2_{\va P})}\,,~~~{\rm where}~~a_i = 8 \pi \ell^2_{\va P} \beta k_i \,.
\ee
The entropy result \eqref{phys2} matches exactly the one obtained in \cite{GP} by exploiting the local CFT symmetry introduced at each puncture by also blowing point particles to infinitesimal, but finite loops. In both cases, such a regularization procedure plays a fundamental role.
 
The presence of the new degeneracy factor \eqref{deg} has been previously postulated in  \cite{GNP} in order to get rid of the quantum gravity correction to the entropy formula associated to a chemical potential term found in \cite{Ghosh-Perez}. A similar analysis then leads to the entropy  
\be\la{entropy}
S=2\pi i \sum_i k_i +o (\sqrt{a_{\va IH}})=\frac{a_{\va IH}}{4 \ell_P^2}+o (\sqrt{a_{\va IH}})\,,
\ee
in agreement with the Bekenstein-Hawking formula for an imaginary B-I parameter. Our result provides yet another evidence, this time originating entirely within the LQG formalism, in support of the new perspective \cite{Complex, Temp} in the LQG black hole entropy calculation allowing for the removal of the numerical restriction on $\beta$ in favor of the physically better motivated analytic continuation to the Ashtekar self-dual connection.

 Notice that, taking the limit $\beta=i,$ the deformation parameter \eqref{q} for plaquettes not containing any puncture reproduces the expression $q=e^{\frac{\pi i }{k}}$ obtained in the Chern-Simons formulation of 2+1 gravity  in presence of a local positive constant curvature ($\Lambda>0$), in agreement with  the deformation parameter of the quantum group $U_q SL(2)$ entering the definition of the Turaev-Viro model that one would expect to appear due to the non-commutativity of the Ashtekar-Barbero connection on $IH$. On the other hand, for plaquettes around the punctures  the deformation parameter \eqref{q} becomes real, again in agreement with the analytic continuation  from $SU(2)$ to $SL(2,\C)$  of the Verlinde formula for the dimension of the Chern-Simons Hilbert space on a punctured 2-sphere performed in \cite{Complex}.

\section{Alternative quantization schemes}\la{sec:alternative}

In this section we want to discuss the relation of the horizon theory and its quantization to certain other results in the literature. On the one hand, the horizon fields $(A,e)$ and the boundary conditions \eqref{BC1} and \eqref{BC2} bear a striking resemblance to the fields and constraints  of 3d  $\Lambda=0$ gravity coupled to particles in a first order formulation.  
On the other hand, in \cite{S} it was suggested to implement the boundary conditions on the horizon field directly in the LQG setting, thereby ignoring the boundary term of the presymplectic structure. The horizon fields $(A,e)$ seem to be ideally suited for this endeavor. Let us discuss these two perspectives in turn. 
\subsection{Connection to 3d gravity with $\Lambda=0$}
3d Euclidean gravity in first order variables, has structure group ISU(2). Generators of this Lie-algebra will be denoted $J_I, P_i, i=1,2,3$ with 
\begin{equation}
\label{eq:isu2}
[P_i,P_j]=0\,, \qquad [P_i,J_j]=\epsilon_{ij}{}^kP_k\,, \qquad [J_i,J_j]=\epsilon_{ij}{}^kJ_k\,. 
\end{equation}
We will closely follow \cite{NP1}.  The gravitational phase space is embedded in the space of ISU(2) connections
\begin{equation}
\mathcal{A}= A^i J_i+e^iP_i
\end{equation}
equipped with Poisson bracket
\begin{equation}
\label{eq:3dgravpoi}
\{ A_{a}^{i}\left(x\right),\, e_{b}^{j}\left(y\right)\} =\epsilon_{ab}\ \delta^{ij}\delta^{(2)}\left(x,\, y\right)\,,
\end{equation}
where $e$ can be thought of as co-triad. Up to a prefactor this is exactly the Poisson structure coming from the boundary presymplectic structure in \eqref{SF3}. 
Coupling of particles to gravity introduces the first class constraints
\begin{equation}
\epsilon^{ab}\mathcal{F}_ab(0)=(p^iJ^i+j^iP_i)\delta^2(x-x_0)
\end{equation}
with the ISU(2) connection $\mathcal{F}= \text{d}_{(\mathcal{A})}\mathcal{A}$ and $p,j$ are the particle dof. Decomposed into translational and SU(2) components:
\begin{equation}\label{C1}
\epsilon^{ab} F_{ab}= p^iJ_i \delta^2(x-x_0)
\end{equation}
with the $SU(2)$ curvature $F=\text{d}_{(A)}A$ and 
\begin{equation}\label{C2}
\epsilon^{ab} (\text{d}_{A}e)_{ab}= j^iP_i \delta^2(x-x_0)\,.
\end{equation}
The Poisson bracket \eqref{eq:3dgravpoi} can be quantized in the standard fashion on L$^{2}(\overline{\mathcal{A}}, \text{d}\mu_{\text{AL}})$ \cite{NP1}. The particle dof obey some constraints on their own and are quantized on the Hilbert space $\mathcal{H}_P=\text{L}^2(\text{SU(2)})$, for details see \cite{NP2}. What is relevant for us are the consequences for the gravitational dof. 
\begin{enumerate}
\item The particles transform under the action of SU(2). Constraint \eqref{C2} implies gauge invariance under the tensor product of gravitational and particle action. This means that particle and gravitational dof have to be coupled by an intertwiner to the trivial representation. 

\item Constraint \eqref{C1} determines the holonomy of loops: In the quantum theory, the connection around a loop is trivial if it does not surround a particle
\begin{equation}
h_\alpha=\mathbb{I}, \qquad \alpha \text{ trivial}
\end{equation}
and is given by an operator on the kinematical Hilbert space of the particle, 
\begin{equation}
\label{eq:hell}
h_{\ell_i}=\Lambda_{i} e^{m_i J_0} \Lambda_{i}^{-1}\,,
\end{equation}
where $m_i$ is a half integer quantum number of the particle, $J_0$ a fixed generator of SU(2) and $\Lambda_{i}$ a multiplication operator on the Hilbert space of the particle. 
\end{enumerate}
Let us compare this to the quantum theory of the isolated horizon. To bring out the analogy, we proceed as in the 3d gravity case, by regarding the boundary conditions \eqref{BC1}, \eqref{BC2}  constraints to be implemented later. Note that due to 
$\text{d}_A e =\beta K\wedge e$
the relation \eqref{KSigma} follows immediately from \eqref{BC2}. Thus the connection $A$ would be commutative initially, as in the BF-formulation of 3d gravity. Hence the kinematical quantization of the gravity dof on the horizon would be the same. Instead of the particle Hilbert space, we would have a piece of the bulk Hilbert space at the puncture
\begin{equation}
\mathcal{H}_p=\bigoplus_{j=0, 1/2,1,\ldots} \mathcal{H}_j\,,
\end{equation}
 which would play the role of the particle Hilbert space. $\mathcal{H}_j$ is the spin $j$ irrep of SU(2).  The consequences of \eqref{BC1}, \eqref{BC2} in the quantum theory are as follows.
Away from the punctures, the quantized version of \eqref{BC2} would enforce gauge invariance of the states on the surface. At the puncture, it would enforce gauge invariance of the tensor product of bulk and boundary state, resulting in kinematical surface states of exactly the same nature  as in 3d gravity with particles. 
We integrate \eqref{BC2} over the disc $D$ bounded by a loop $\ell$ surrounding a puncture:
\begin{equation}
\int_D h_{x0}^{-1}\, F(x)\, h_{x0}\ \text{d}^2x= c \int_D h_{x0}^{-1} \,\Sigma(x)\, h_{x0}\ \text{d}^2x. 
\end{equation}
Here $h$ are holonomies from some fixed point $0$ on $\ell$ to the point $x$ and c is a constant. 
The quantization of the right-hand side is given by $h^{-1}_{p0}\,L^iJ_i\,h_{p0}$, where $L$ is an angular momentum operator on $\mathcal{H}_p$, and it can be shown that there is a basis such that $L^iJ_i$
is $\lambda J_0$ for some fixed generator $J_0$ of SU(2) and suitable numbers $\lambda$, i.e. 
\begin{equation}
\label{eq:f1}
\int_D h_{x0}^{-1}\, F(x)\, h_{x0}\ \text{d}^2x= c\lambda h_{p0}^{-1}\,J_0\, h_{p0}\,.
\end{equation}
On the other hand, expanding \eqref{eq:hell} up to first order, we have 
\begin{equation}
\label{eq:f2}
\mathbb{I}+\int_D h_{x0}^{-1}\, F(x)\, h_{x0}\ \text{d}^2x =\mathbb{I} + \Lambda m J_0 \Lambda^{-1}\,.
\end{equation}
Comparing \eqref{eq:f1} and \eqref{eq:f2} we see a strong similarity in the quantum theory. This suggests that the boundary theory might alternatively be quantized as Euclidean $\Lambda=0$ 3d gravity
with particles. 
A quantization of the exponential of \eqref{BC1} has been suggested in \cite{ST}, leading to the expectation value
\begin{equation}
\text{tr}\; h_\ell = \frac{q^{(2j+1)}-q^{-(2j+1)}}{q-q^{-1}}
\end{equation}
for a given quantum number $j$ of the puncture (including $j=0$ for the case of a loop not enclosing any puncture). This is of the same form as \eqref{q-dim-j} (up to the sign of $q$), however,  $q$ is now given by
$q= e^{\frac{\pi i}{k}}$,
with the level $k$ from \eqref{eq:k}. That $q$ is different is no contradiction to \eqref{q-dim-j}, as the holonomies belong to different connections. Moreover, as we pointed out at the end of Section \ref{sec:entropy}, once we take $\beta=i$ the deformation parameter \eqref{q} reproduces the usual expression above obtained in the literature.

Let us conclude with an observation on a possible description in terms of bulk operators.
In \cite{S} it was suggested that the boundary quantum theory could be taken as a restriction to the boundary of the bulk quantum theory. 
\begin{figure}[]
\centering
  \includegraphics[width=5.cm]{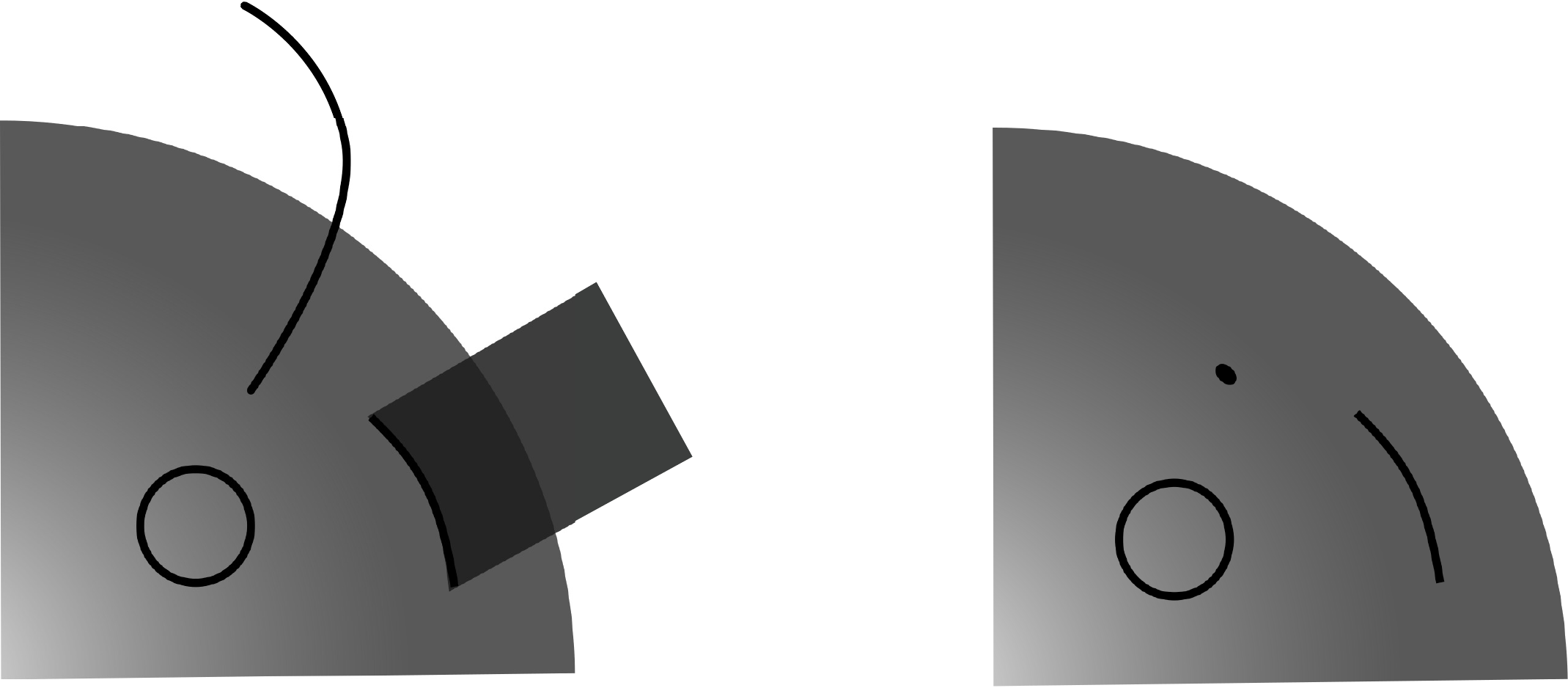}\,.
   \caption{Operators of the bulk holonomy-flux algebra give operators in the surface theory.}
\label{fi:surface}
\end{figure}
Indeed, this could be carried out here. Holonomies entirely in the surface represent the connection on the surface, holonomies that end on the surface contribute defects or particles, see Figure \ref{fi:surface}. The flux operators in the bulk, which have a transversal intersection with the surface, give, when restricted to this intersection, an operator that has the same commutation relations with holonomies as the two-dimensional flux \eqref{flux}. This is remarkable, because they do correspond to different classical quantities. There is no contradiction, however, because the operators are the results of quantizing two different presymplectic spaces, the bulk and the surface one, respectively.
Constraint \eqref{BC1} has been investigated in this context \cite{ST}, but it is not yet clear wether solutions are proper states on the holonomy flux algebra. The constraint \eqref{BC2} has not yet been investigated in this context. Presumably, it is again linked to gauge invariance on the horizon.

\section{Summary}\la{sec:summary}
As a first step of our analysis, we have re-expressed the IH conserved presymplectic form in terms of first order 2+1 gravity variables. That this is possible is very satisfying, as these variables have an immediate geometric interpretation. 
Then we have  quantized the boundary theory uniquely in terms of LQG techniques, without relying on  structures of the Chern-Simons theory on a punctured 2-sphere. 
We have shown how the physical scalar product of canonical LQG can be used to compute the quantum IH state degeneracy, leading to an entropy in agreement with the Bekesntein-Hawking formula for $\beta=i$. 

Our analysis avoids several ambiguities present in the usual coupling of the bulk LQG and the boundary Chern-Simon Hilbert spaces performed in previous literature, thus presenting a more coherent and sound picture of black hole entropy calculation in LQG, based on a nice interplay between the three and the four-dimensional theories. 

We would like to point out again how a key ingredient for the entropy result, namely the new degeneracy factor \eqref{deg}, is represented by the punctures regularization via finite circles, which appears naturally when employing LQG structures. It would be interesting to study the algebra of observables that one could define living on these new boundaries, in analogy with the analysis of \cite{GP}, and investigate if a Virasoro structure might emerge  just from within the LQG formalism. 

Finally, we note that the horizon punctures behave very much like particles in the 2+1 quantum gravity describing the horizon. This beautiful picture underscores again that the horizon thermodynamics is the thermodynamics of these punctures.

\end{document}